\documentclass{article} 

\usepackage[ruled]{algorithm2e}
\usepackage{graphicx}
\usepackage{amsmath}
\usepackage{bm}
\usepackage{rotating}
\usepackage{amssymb}
\usepackage{url}

\usepackage{amsthm}

\SetAlFnt{\small}
\SetAlCapFnt{\small}
\SetAlCapNameFnt{\small}
\SetAlCapHSkip{0pt}
\IncMargin{-\parindent}

\begin{document}


\title{Modelling Word Burstiness in Natural Language \\ ~ \\  \large A Generalised P\'olya Process for Document Language Models in Information Retrieval}
\author{Ronan Cummins}
\date{}
\maketitle

\begin{abstract}

We introduce a generalised multivariate P\'olya process for document language modelling. The framework outlined here generalises a number of statistical language models used in information retrieval for modelling document generation. In particular, we show that the choice of replacement matrix ${\bf M}$ ultimately defines the type of random process and therefore defines a particular type of document language model. 
We show that a particular variant of the general model is useful for modelling term-specific burstiness. Furthermore, via experimentation we show that this variant significantly improves retrieval effectiveness on a number of small test collections.

\end{abstract}

\section{Introduction}

Document language modelling is a crucial component in statistical approaches to information retrieval \cite{ponte98,hiemstra2001,zhai04}. These types of approach are generative in nature, in so far as they aim to estimate a model of the document generation process. Often these generative document models are simple unigram mixture models that incorporate information from the background collection as well as information from a specific document. Thus, the main challenge is one of parameter estimation and consequently various smoothing approaches have been studied \cite{zhai04} that aim to better estimate the parameters of the document models given a collection of documents. In information retrieval, once the document models have been estimated, documents can be ranked by the likelihood of their document model generating the query (i.e. the query-likelihood approach \cite{ponte98}).

%

Recently, document language models that exhibit a self-reinforcing property, via a multivariate P\'olya process, have been shown to significantly increase retrieval effectiveness \cite{cummins15a}. This approach captures word burstiness in a document-specific way. In this paper, we take this approach further and outline a more general process for statistical document modelling, one which encompasses a number of existing document language models as well as a number of novel variants. 

In Section~2, we outline a general model of document generation in terms of an \emph{urn} scheme which crucially models the dynamics of document generation using a matrix ${\bf M}$. Section 3 discusses specific instantiations of ${\bf M}$, ones which lead to determining different generative distributions. Section 4 outlines how the new general model is used for document retrieval. Section 5 outlines our experimental set-up. The results of a number of experiments are reported in Section 6. Finally, in Section 7 we conclude with a discussion and outline some future work.

\section{Generalised P\'olya}

Let ${\bf u}_0$ be an urn initially containing $|{\bf u}_0|$ balls in total where each ball is one of $v$ distinct colours. Starting at time $i=0$, a ball is drawn with replacement from the urn and a number of additional balls (possibly of different colours) are added\footnote{such that the mass of the urn never decreases} to the urn according to a replacement matrix ${\bf M}$. Each row of this matrix determines the number of additional balls of each colour to add to the urn and the row is selected according to the colour of the ball drawn from the urn. Therefore, the dynamical nature of this random process is defined by this $v^2$ matrix ${\bf M}$. 

Now, if $d = \{t_i\}$ is a sequence of observations indexed from $i=0$ to $i=|d|-1$ drawn from the urn, then the state of the urn can be describe as a recurrent process as follows:

\begin{equation}
{\bf u}_{i+1}  = {\bf u}_{i} + {\bf e}_{t_i} \cdot {\bf M}
\end{equation}
where ${\bf e}_{t_i}$ is a standard basis vector (a.k.a a one-hot vector\footnote{a vector that is a $1$ in dimension $t_i$ and $0$ elsewhere}). At a particular time $i$ and for an observation $t_i$, the process essentially selects row $t_i$ of the matrix ${\bf M}$ and then combines it with the $i^{th}$ state of the urn ${\bf u}_{i}$. This imbues the urn with a reinforcing property, and in specific cases results in the traditional P\'olya urn scheme. The generalised P\'olya process is thus defined by both the replacement matrix ${\bf M}$ and the initial parameters ${\bf u}_0$ and so the likelihood of seeing a specific sample $d$ can be written as $p(d|{\bf u}_0, {\bf M})$. One can interpret ${\bf u}_0$ as completely defining the probability of seeing a particular coloured ball on the first draw, and can interpret ${\bf M}$ as defining the dynamics of the urn. These types of model are well-suited to modelling natural language where different coloured balls represent different word-types \cite{goldwater11} and where the number of balls of each colour in the urn is proportional to the probability of generating a specific word. 

\section{Choices of ${\bf M}$}

In this section, we outline four different variants of ${\bf M}$. The first two of these variants do not rely on estimating ${\bf M}$ from data and already appear in the literature \cite{zhai01,cummins15a} in one form or another. They have also been implemented and have resulted in increases in retrieval effectiveness and in advances of a theoretical nature. The latter two models have yet to be realised.

\subsection{Zero Matrix}

If we choose ${\bf M}$ to be the zero (denoted ${\bf 0}$) matrix, the initial state of the urn does not change during the process. As the drawing of a particular colour does not effect subsequent draws, choosing the zero matrix is equivalent to using a multinomial language model.  

\begin{equation}
{\bf M} =
 \begin{pmatrix}
  0 & 0 & \cdots & 0 \\
  0 & 0 & \cdots & 0 \\
  \vdots  & \vdots  & \ddots & \vdots  \\
  0 & 0 & \cdots & 0
 \end{pmatrix}
\end{equation}
Estimating the maximum-likelihood parameters of ${\bf u}_{0}$ is trivial as they have closed-form solutions. Using this multinomial document language model (with different types of smoothing) is quite effective for retrieval \cite{zhai04}. In particular, it was shown that a document language model using Dirichlet prior smoothing is particularly effective.

\subsection{Identity Matrix}

If we choose ${\bf M}$ to be the identity matrix (denoted ${\bf 1}$), the state of the initial urn changes with a self-reinforcing property, as the drawing of a particular coloured ball only reinforces the urn with another copy of that particular colour. This process is equivalent to the multivariate P\'olya urn or Dirichlet-compound multinomial (DCM) and has been shown to generate text according to the power-law characteristics of natural language.  

\begin{equation}
{\bf M} =
 \begin{pmatrix}
  1 & 0 & \cdots & 0 \\
  0 & 1 & \cdots & 0 \\
  \vdots  & \vdots  & 1 & \vdots  \\
  0 & 0 & 0 & 1
 \end{pmatrix}
\end{equation}
This document language model has been implemented recently and has shown significant increases in retrieval effectiveness over the multinomial model \cite{cummins15a}. Furthermore, it has led to a number of theoretically interesting properties. It models word burstiness in a document-specific manner which in turn has led to a better understanding of both the scope and verbosity hypotheses \cite{cummins16}. This distribution contains only one extra parameter compared to the multinomial (a parameter which models the burstiness of terms). However, one of the weaknesses of the model is that is assumes that all terms are equally bursty. 

\subsection{Diagonal Matrix}

If we choose ${\bf M}$ to be any positive diagonal matrix, the state of the initial urn again changes with a self-reinforcing property, but this time the amount of reinforcement is different depending on the colour of the ball drawn. This process can captures term-specific burstiness where certain words are more likely to repeat within a specific sample than others.  

\begin{equation}
{\bf M} =
 \begin{pmatrix}
  m_{1,1} & 0 & \cdots & 0 \\
  0 & m_{2,2} & \cdots & 0 \\
  \vdots  & \vdots  & m_{v-1,v-1} & \vdots  \\
  0 & 0 & 0 & m_{v,v}
 \end{pmatrix}
\end{equation}
In this case, we need to determine the $v$ parameters in this matrix. In this model, each term has an initial probability of occurring but also a specific parameter controlling its re-occurrence. We can set these extra parameters to some intuitively motivated values or can estimate them directly from data. The latter can be done using numerical optimisation or sampling methods. This is the model that is the focus of the experiments in this paper.

\subsection{Full Replacement Matrix}

If we choose ${\bf M}$ to be any matrix with the constraint such that for any replacement row vector ${\bf m}_k$ then $\sum^{v}_{j=1} {\bf m}_{kj} >0$, the state of the urn can change in all $v$ dimensions for any draw. The constraint is necessary to ensure that the process can continue ad infinitum (i.e. the urn is never left with less mass than previous). 

\begin{equation}
{\bf M}_{|v|,|v|} =
 \begin{pmatrix}
  m_{1,1} & m_{1,2} & \cdots & \cdots \\
  m_{2,1} & m_{2,2} & \cdots & \vdots \\
  \vdots  & \vdots  & \vdots & \vdots  \\
  m_{v,1} & 0 & 0 & m_{v,v}
 \end{pmatrix}
\end{equation}
This model can capture dependencies between different word-types within documents. For example, it may be that the words \emph{dna} and \emph{blood} tend to occur in the same documents. While this model is theoretically interesting, the remainder of this paper is focussed on the model presented in Section~3.3.

\subsection{Discussion}

The general model outlined here (Eq.~1) is an intuitive statistical generative model of documents. The vector ${\bf u}_i$ can be seen as storing the state of the model at a particular time $i$. Both the multinomial and multivariate P\'olya urn (SPUD \cite{cummins15a}) language model are specific instances of this model and are instantiated by different settings of ${\bf M}$. Given that the SPUD language model significantly improves upon the multinomial model in information retrieval, the further extensions hold the promise of improved performance and of greater theoretical understanding. Furthermore, it is worth noting that the dependencies that the models\footnote{those from from Section~3.2 onwards} capture, span a greater distance than $n$-gram models (i.e. a word occurring at the start of a document affects the choice of word at the end of a document).

The main challenges to implementing the remaining model variants are in estimating ${\bf M}$ and ${\bf u}_0$ from a large background model (document collection ${\bf D}$) and subsequently in inferring the initial state of each document model. For large scale collections this a computationally expensive inverse problem. However, the upcoming section will outline some promising initial experiments with regard to the third variant of the general model (i.e. modelling term-specific burstiness for retrieval). The fourth variant is left for future work.

\section{Parameter Estimation}

This section is concerned with estimating from data the parameters of the general background language model ${\bf u}_{0}$ and ${\bf M}$, and the unsmoothed document model ${\bf u}_{0}^d$.

\subsection{Background Model Estimation}

In the language modelling approach to information retrieval, it is common to estimate a background collection model using all of the documents $d$ from the collection $D$. Therefore, we first estimate the parameters of the background model (${\bf u}_{0}$ and ${\bf M}$) that has generated all the documents in the collection $D$. We adopt a Bayesian approach to this problem and aim use the posterior mean as the point-estimate as follows: 
\begin{equation}
\mathbb{E}({\bf u}_{0}, {\bf M}) = \int ({\bf u}_{0}, {\bf M}) \cdot p({\bf u}_{0}, {\bf M} | D )\cdot  d_{{\bf u}_{0}, {\bf M}}
\end{equation}
In general, estimating these parameters is computationally expensive for large scale document collections. In this paper, we use smaller collections and make use of MCMC sampling to approximate the posterior distribution. The main bottleneck in such an approach is estimating the likelihood of the data given the parameters (for many parameter samples). However, there are alternative approaches to setting these parameters. For example, if ${\bf M}$ is set to a fixed value, then we only need to estimate ${\bf u}_{0}$ from data. We will outline the specifics of the estimation approach in Section~5.

\subsection{Document Model Estimation}

Once ${\bf M}$ is known\footnote{via estimation or intuition}, the parameters of the unsmoothed document model $({\bf u}_{0}^d)$ that generated a specific document $d$ need to be estimated. Again we take a Bayesian approach and estimate the expectation of the posterior as follows:

\begin{equation}
\mathbb{E}({\bf u}_{0}^d) = \int  {\bf u}_0^d \cdot p({\bf u}_0^d|{\bf M},d) \cdot d_{{\bf u}_0^d}
\end{equation}
where it is worth noting that these parameters need to be estimated for each document $d$ in the collection $D$. ${\bf M}$ is fixed here as it represents the general dynamics of document generation (not of a specific document). Once both ${\bf u}_{0}$ and ${\bf u}_{0}^d$ are estimated, they can be linearly smoothed using a single hyperparameter as follows: 

\begin{equation}
({\bf u}_{0}^{d'})  = ((1-\omega)\cdot {\bf u}_{0}^d + \omega \cdot {\bf u}_{0})
\end{equation}
where $0 \leq \omega \leq 1$ is a tuning parameter and ${\bf u}_{0}^{d'}$ is the final document model. The parameters of ${\bf u}_{0}^{d'}$ can be interpreted as the initial proportions of words of each type in the document model before any draws have been made.

\subsection{Query-Likelihood for Retrieval}

We adopt the well-known query likelihood approach. Therefore, each document $d$ can be ranked by determining the probability of their document model generating the query as follows:

\begin{equation}
p(q| {\bf u}_{0}^{d'}, {\bf M}_q)
\end{equation}
where ${\bf M}_q$ is a dynamical model for query generation. It may be the same as ${\bf M}$ (for documents) or could be something as simple as the zero matrix. In this paper, we assume a zero matrix where is no query dynamics. We justify this by noting that queries are typically much shorter than documents and are motivated by a very different need when compared to documents. Therefore, in this paper we rank documents according to the following formula once both ${\bf u}_{0}$ and ${\bf u}_{0}^d$ are estimated: 

\begin{equation}
log~p(q|{\bf u}_{0}^{d'}, {\bf 0}) = \sum_{t \in q} log ( \frac{\mu_d}{\mu_d + \mu } \cdot \frac{u_{0_t}^d}{|{\bf u}_0^d|} + \frac{\mu}{\mu_d + \mu } \frac{u_{0_t}}{|{\bf u}_0|} ) 
\end{equation}
where ${|{\bf u}_0^d|}$ and ${|{\bf u}_0|}$ are the mass of the document model and background model used to calculate actual probabilities. However, when estimating the parameters of ${{\bf u}_0^d}$ of a single document $d$, the mass (i.e. ${|{\bf u}_0^d|}$) is often not constrained. Therefore, we introduce $\mu_d$ as the initial mass of the document model and set it to the number of unique terms in $d$. Furthermore, by re-writing the retrieval formula, we have subsumed $\omega$ into $\mu$, thus leaving $\mu$ the only free hyperparameter in the model.

\section{Experiments}

The aim of our experiments is to test the third variant (denoted GSPUD from now on) of the general model and compare it to the models when ${\bf M=0}$ (MULT) and ${\bf M=1}$ (DCM). This third model contains $v$ extra parameters compared to the multivariate P\'olya scheme (DCM). These extra parameters model the burstiness of each term individually. We outline two approaches to finding them. One method involves setting these parameters according to some heuristic (i.e. intuition), and the second method involves estimating the parameters directly from the data using numerical methods. Before this we first outline our use of Metropolis-Hastings MCMC sampling for estimating the parameters (${\bf u}_0$ and ${\bf u}_0^d$) of the models outlined in this work.

\subsection{Estimation}
For all of the models implemented in this work, we make use of MCMC sampling to estimate the posterior expectation of the parameter distribution. We use the Metropolis-Hastings algorithm as it is easily implemented (i.e. one only needs to be able to calculate the likelihood of the data given a set of parameters), can deal with complex distributions (i.e. we do not know the conjugate distribution of the generalised P\'olya distribution), and can deal with high-dimensional data (i.e. the background model will contain thousands of parameters/words). Other techniques may indeed prove to be faster, but for our purposes we only need a method of arriving at suitable estimates. 

In particular, we use the Metropolis-Hastings algorithm with a Gaussian proposal distribution (variance of 0.01 for the background model and variance of 0.25 for each document model). For background models, we run the chain for 500,000 samples discarding the first 50,000 samples (i.e. burn-in period). For individual document models, we run the chain for 200,000 samples discarding the first 20,000 samples. We estimate the expectation of the posterior parameter distribution using the likelihood function with a uniform prior. We start the algorithm with a uniform distribution (all parameters are set to $1.0$).\footnote{In practice we sample in the log parameter space and then exponentiate to avoid negative values which are invalid for these models.} We used these techniques on all variants of our models. We used this method for the multinomial model in order to compare the effectiveness of the sampling algorithm on models where we have closed-form maximum-likelihood estimates. This was especially useful during development.\footnote{Code available at \url{https://github.com/ronancummins/gen_polya}}

\subsection{Setting ${\bf M}$ Heuristically}

One method of setting the $v$ parameters of ${\bf M}$ is to set them according to a heuristic. One measure of term-burstiness that has been outlined in the literature is as follows:

\begin{equation}
m_{t,t} = bs_t = \frac{cf_t}{dt_t}
\end{equation}
where $cf_t$ is the frequency of the term in the collection, and $df_t$ is the document frequency of the term. This quantity measures the average frequency of a word in a document given that it has occurred once. The measure has appeared extensively in the information retrieval literature \cite{kwok96,franz00,cummins06}. It has a lower bound of $1.0$ (which in fact is the default value of term-burstiness in the multivariate P\'olya scheme outlined in Section~3.2). Once ${\bf M}$ is set in this way, we can estimate the initial parameters ${\bf u}_0$ and ${\bf u}_0^d$ using MCMC.

\subsection{Data}

Due to the computationally expensive nature of the task, we use three small test collections (Medline, Cranfield, and CISI)\footnote{Available from \url{http://ir.dcs.gla.ac.uk/resources/test_collections/}}. Although, these collections are rather old and are quite small, when compared to modern large scale Web collections such as ClueWeb, they can provide some insights and intuitions into what natural language characteristics are being captured by the new model. We removed standard stopwords and stemmed the documents and queries. Table~\ref{tab:collections} shows some of the characteristics of the collections after preprocessing. 

\begin{table*}[!ht] 
\centering
\small
{\renewcommand{\arraystretch}{1.0}
\begin{tabular}{|  r || r | r | r | r |}

\hline
Collection		&	Medline	& Cranfield	& CISI	\\
\hline
\# docs			&	1,033		& 1,400		& 1,460			\\
\# vocab ($v$) 	& 	8,764		& 5,769		& 7,062			\\														
\# tokens 		& 	97,175		& 153,276	& 115,527		\\
\# qrys 		& 	30			& 225		& 76			\\
\hline

\end{tabular}}
\caption{Test Collections Details}
\label{tab:collections}
\end{table*}

\section{Results}
In this section we report our results along with some qualitative analysis that aim to interpret them further.

\subsection{Background Estimation}

First, in Table~\ref{tab:perplexity} we look at how well our different models fit the background collection. We use perplexity to compare the performance of our model on the data on which the models were trained. Perplexity is a measure of how surprised our model is by the data (a lower perplexity indicating a better model). The number of parameters for each model is shown in the table, where the vocabulary ($v$) can be found in Table~\ref{tab:collections}. 

As a benchmark for our sampling algorithm, we used MCMC to estimate the parameters of the multinomial model (${\bf M = 0}$). We see that our MCMC estimates (MULT$_{mc}$) result in a model which has a very similar perplexity to the maximum-likelihood estimates (MULT$_{mle}$). Furthermore, the DCM$_{mc}$ model (${\bf M = 1}$) improves substantially over the multinomial model with only one extra parameter. The generalised P\'olya model with $v$ burstiness parameters set to $bs_t$ (GSPUD$_{bs_{t}}$) improves over the DCM model. Finally, the generalised P\'olya model with $v$ burstiness parameters estimated via MCMC (GSPUD$_{mc}$) has the lowest perplexity of all models. While this tells us that the parameters are modelling aspects of the document collection such that they are better able to predict the sequence of words, it is unclear whether these results extend beyond the collections on which they were trained. While these experiments provide a useful sanity check regarding the ability of our models to better fit to data, we need to use these models in a retrieval setting to ultimately determine if they can improve effectiveness.

\begin{table*}[!ht] 

\centering
\small
{\renewcommand{\arraystretch}{1.0}
\begin{tabular}{|  l || c | c || c | r | r | r |}
\hline
					& \multicolumn{2}{|c||}{Estimation}	& 		&	\multicolumn{3}{|c|}{Datasets}	\\
\hline
Model				&${\bf u_0}$	& ${\bf M}$	& 	\# params	&	Medline	& Cranfield	& CISI	\\
\hline	
MULT$_{mle}$		& mle			& ${\bf 0}$			& $v-1$		& 2047.4		& 971.6	& 1309.7		\\														
MULT$_{mc}$			& mcmc 			& ${\bf 0}$			& $v-1$		& 2079.2		& 980.2	& 1325.0		\\
DCM$_{mc}$			& mcmc 			& ${\bf 1}$			& $v$		& 1728.8		& 883.1	& 1282.2		\\
GSPUD$_{bs_{t}}$	& mcmc  		& $bs_t$			& $2v$		& 1369.1		& 688.4	& 1248.4		\\
GSPUD$_{mc}$		& mcmc 			& mcmc				& $2v$		& 1152.7		& 597.8	& 999.6			\\

\hline

\end{tabular}}
\caption{Perplexity (nats) of different language models trained on the background corpus.}
\label{tab:perplexity}
\end{table*}

\subsection{Document Models and Retrieval Performance}
\begin{figure}[!ht] 
\begin{center}
\begin{tabular}{c c c}
   	\includegraphics[height=3.1cm,width=3.7cm]{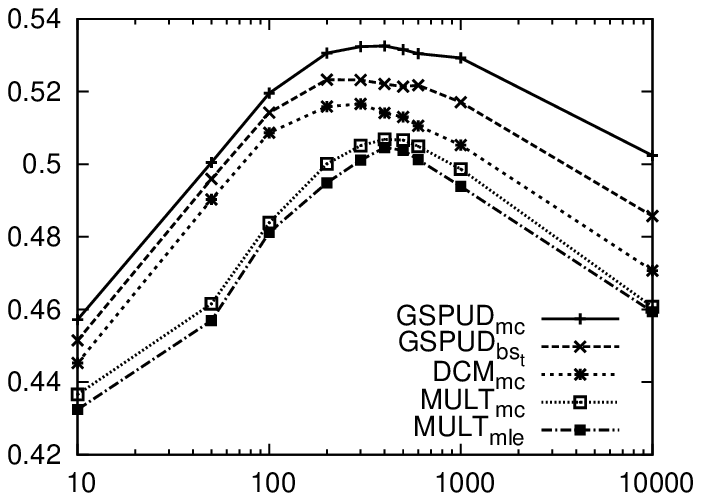} &    
	\includegraphics[height=3.1cm,width=3.7cm]{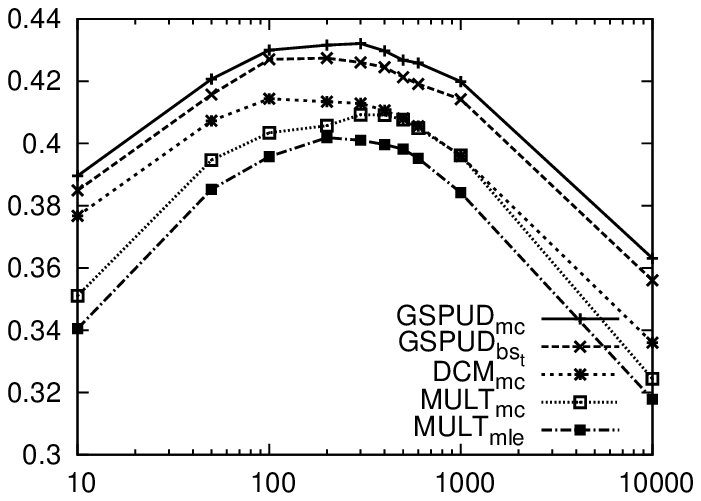} & 
	\includegraphics[height=3.1cm,width=3.7cm]{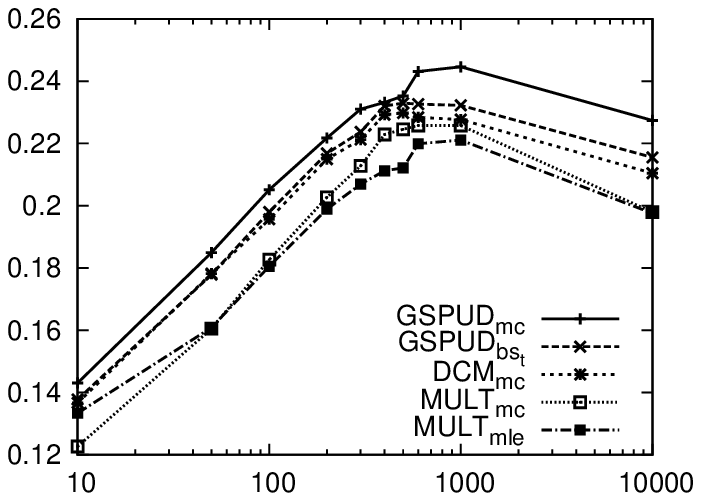} \\
\end{tabular}
\caption{MAP of different models for varying values of $\mu$ Medline, Cranfield, and CISI.}
\label{fig:tuning}
\end{center}
\end{figure}

Table~\ref{tab:retrieval} shows the optimal performance of each model when $\mu$ is tuned per collection over the values
$\mu = {10,50,100,200,300,400,500,1000,10000}$. Fig.~\ref{fig:tuning} shows the trends when $\mu$ varies. In all cases we can see that the GSPUD models outperform both the DCM and MULT models. The results are statistically significant. The best performing model is the GSPUD model that estimates its burstiness parameters directly from data rather then heuristically. Interestingly, the 
MULT$_{mc}$ approach outperforms MULT$_{mle}$ suggesting that Bayesian estimates might be better than 
maximum-likelihood estimates for the retrieval task.

\begin{table*}[!ht] 

\centering
\small
{\renewcommand{\arraystretch}{1.0}
\begin{tabular}{|  l || c | c ||  l | l | l |}
\hline
					& \multicolumn{2}{|c||}{Estimation}	 		&	\multicolumn{3}{|c|}{Datasets}	\\
\hline
Model				&${\bf u_0},{\bf u_0^d}$	& ${\bf M}$	& 			Medline	& Cranfield	& CISI		\\
\hline	
MULT$_{mle}$			& mle			& ${\bf 0}$				& 0.504				& 0.402			& 0.221				\\														
MULT$_{mc}$				& mcmc 			& ${\bf 0}$				& 0.506				& 0.409			& 0.225				\\
DCM$_{mc}$				& mcmc 			& ${\bf 1}$				& 0.517$^{m}$		& 0.414$^{m}$	& 0.230$^{m}$		\\
GSPUD$_{bs_{t}}$		& mcmc  		& $bs_t$				& 0.523$^{md}$		& 0.427$^{md}$	& 0.233$^{md}$		\\
GSPUD$_{mc}$			& mcmc 			& mcmc					& 0.533$^{md}_{g}$	& 0.432$^{md}_{g}$	& 0.245$^{md}_{g}$		\\

\hline

\end{tabular}}
\caption{Performance (Mean Average Precision) of different language models with $\mu$ tuned per collection. The keys $m,d,g$ mean that the result is statistically significant compared to MULT, DCM, and GSPUD$_{bs_{t}}$ respectively at the $p<0.01$ level using a permutation test.}
\label{tab:retrieval}
\end{table*}

\subsection{Qualitative Analysis}

In order to qualitatively evaluate what our models are learning, we now analyse a number of words that appear in the Medline collection. The three words \emph{also}, \emph{dna}, and \emph{refer} are shown with some of their collection statistics. In a multinomial model, \emph{also} and \emph{dna} would have a very similar discriminative value because they occur nearly the same number of times throughout the corpus. Models that only use a type of $idf_t$ would assign a similar discriminative value to \emph{refer} and \emph{dna}. However, it is intuitive that \emph{dna} is a much more useful word in general due to its burstiness. We can see that the burstiness of \emph{dna} is much higher than the other terms when estimated from data (i.e. $\hat{m}_{t,t}$). It is worth remembering that in the GSPUD model, each word is characterised by two parameters, one that controls the probability of being drawn from the initial urn at time $t=0$ and one that controls the burstiness. For the GSPUD model, the initial probability of a word being drawn from the urn is very closely correlated with the document frequency and is close to the probability $\frac{df_t}{\sum_{t'} df_{t'}}$. The ideas of word burstiness have been around for quite a while with a number of interesting papers written by Church and Gale \cite{church95,church99} outlining the properties of \emph{important} keywords in documents.

\begin{table*}[!ht] 

\centering
\small
{\renewcommand{\arraystretch}{1.0}
\begin{tabular}{|  l || r | r | r || r |  r |}
\hline
term ($t$)			&	$cf_t$	& $df_t$	& 	$bs_t$	& $ \hat{u}_{0_t}$		& $\hat{m}_{t,t}$	\\
\hline	
also 				&  216	& 180			& 1.20		& 37.86		& 10.22		\\
dna					&  214	& 47			& 4.55		& 9.28		& 266.27	\\
refer				&  51	& 47			& 1.08		& 9.08		& 3.22		\\

\hline

\end{tabular}}
\caption{Some statistics and example estimates of the parameters of the GSPUD$_{mc}$ for three words}
\label{tab:examples}
\end{table*}

\section{Summary}
This paper has introduced a family of statistical language models inspired by a classic urn model. We have shown that it is the replacement matrix ${\bf M}$ that defines the dynamics of the model. We have implemented a variant of the model which models burstiness in a term-specific manner. We have shown that the parameters of the model can be estimated from data using sampling techniques. 

Furthermore, we have incorporated the new language model into a retrieval framework and shown that retrieval effectiveness improves significantly over a highly competitive baseline language model. Although our experiments are conducted on small test collections (because parameter estimation is computationally expensive), the results are promising. We believe that this is first paper that deals with term-specific burstiness in such a principled probabilistic manner.

\bibliographystyle{plain}

\end{document}